\documentclass[a4paper]{spie}  

\usepackage[pdftex]{graphicx}

\title{The in-flight spectroscopic performance of the Swift XRT CCD camera
  during 2006-2007 }

\author{O. Godet\supit{a}, A. P. Beardmore\supit{a}, A. F. Abbey\supit{a}, J.
  P.  Osborne\supit{a}, K. L. Page\supit{a}, L. Tyler\supit{a}, D. N.
  Burrows\supit{c}, P.  Evans\supit{a}, R. Starling\supit{a}, A. A.
  Wells\supit{a}, L.  Angelini\supit{b}, S. Campana\supit{d}, G.
  Chincarini\supit{d,e}, O.  Citterio\supit{d}, G.  Cusumano\supit{f}, P.
  Giommi\supit{g}, J. E.  Hill\supit{b}, J.  Kennea\supit{c}, V.
  LaParola\supit{f}, V.  Mangano\supit{f}, T. Mineo\supit{f}, A.
  Moretti\supit{d}, J. A.  Nousek\supit{c}, C.  Pagani\supit{c}, M.
  Perri\supit{g}, M. Capalbi\supit{g}, P. Romano\supit{d,e}, G.
  Tagliaferri\supit{d}, F.  Tamburelli\supit{g} \skiplinehalf
  \supit{a}University of Leicester, University Road, Leicester, LE1 7RH, UK; \\
  \supit{b}NASA-GSFC, Greenbelt, MD 20771, USA;\\
  \supit{c}Pennsylvania State University, 525 Davey Lab, University Park, PA 16802, USA;\\
  \supit{d}INAF-Osservatorio Astronomico di Brera, Via E. Bianchi 46, 23807,
  Merate, LC, Italy;\\
  \supit{e} Universit\`a degli Studi di Milano, Bicocca, Piazza delle Scienze
  3, I-20126, Milano, Italy;\\
  \supit{f}INAF-IASF, Via U. La Malfa 153, 90146 Palermo, Italy;\\
  \supit{g}ASI-ASDC, Via G. Galilei, I-00044 Frascati, Italy }

\authorinfo{Further author information: (Send correspondence to
  O.G.)\\O.G.: E-mail: og19@star.le.ac.uk, Telephone: +44 116 223 1039\\
  A.P.B.: E-mail: apb@star.le.ac.uk, Telephone: +44 116 252 3583}

\begin{document} 
\maketitle 

\begin{abstract}
  
  The Swift X-ray Telescope focal plane camera is a front-illuminated MOS CCD,
  providing a spectral response kernel of 135 eV FWHM at 5.9 keV as measured
  before launch. We describe the CCD calibration program based on celestial
  and on-board calibration sources, relevant in-flight experiences, and
  developments in the CCD response model. We illustrate how the revised
  response model describes the calibration sources well.  Comparison of
  observed spectra with models folded through the instrument response produces
  negative residuals around and below the Oxygen edge. We discuss several
  possible causes for such residuals. Traps created by proton damage on the
  CCD increase the charge transfer inefficiency (CTI) over time. We describe
  the evolution of the CTI since the launch and its effect on the CCD spectral
  resolution and the gain.
\end{abstract}




\section{INTRODUCTION}

\label{sec:intro}  

The Swift gamma-ray burst satellite~\cite{gehrels} was successfully launched
on 2004 November 20. Since then, it has provided observations and positions of
GRBs and their afterglows to observers and robotic telescopes typically within
a minute, thanks to its three instruments: the wide-field Burst Alert
Telescope~\cite{Barthelmy} and the two narrow-field instruments, X-Ray
Telescope~\cite{Burrows} and UV/Optical Telescope~\cite{Roming}\,. When a GRB
is detected by the BAT and a slew is possible, Swift automatically re-points
to bring the burst within the field of view of the XRT and the UVOT.

The XRT uses a grazing incidence Wolter-1 telescope consisting of a thermally
controlled carbon fibre telescope tube, an X-ray mirror system of 12
concentric gold-coated electroformed Ni shells with a 3.5m focal length, and a
Focal Plane Camera Assembly (FPCA) housing an e2v CCD-22 with $600\times 602$
image pixels, located behind an optical blocking filter with an optical
transmission of about $0.25\%$. The CCD is mounted on a thermoelectric cooler
connected via a heat pipe to an external radiator. The FPCA also includes an
autonomous Sun shutter, four $^{55}$Fe calibration sources and a substantial
mass of Al proton shielding (which also reduces thermal variations).  To avoid
the effects of pile-up, the XRT is able to autonomously select one of the
three following readout modes~\cite{Hill} according to the source brightness:
Photo-diode (PD) mode at highest count rates with a 0.14 ms time resolution
and no spatial information; Windowed Timing (WT) mode at moderate count rates
with a 1.8 ms time resolution and 1-D spatial information; at lower count
rates Photon Counting mode (PC) with a 2.5 s time resolution and 2-D spatial
information.

During the first year in orbit two major incidents occurred which required
modification of the instrument's operation, although they have not affected
its scientific productivity. First, before the CCD was cooled to its nominal
operating temperature of $-100^{\circ}$C, the XRT thermo-cooler (TEC) power
supply system apparently failed, and therefore the XRT has to rely on passive
cooling via the heat pipe and radiator in combination with enhanced management
of the spacecraft orientation to reduce the radiator view of the sunlit earth.
In flight the XRT is nowadays operated with CCD temperatures of -75 to
-52$^{\circ}$C (see Kennea et al.~\cite{kennea} for more details).  Secondly,
on 2005 May 27 the XRT was hit by a particle (micro-meteoroid) which scattered
off the mirror system to hit several CCD pixels, causing new bright pixels,
one bright column and two bright column segments~\cite{Abbey}\,.  Charge
leakage from the top of the bright column affects its immediate neighbours.
The evolution of charge leakage depends on the CCD temperature which can now
only be controlled by orientation of the spacecraft.  After this event, the
optical filter showed no sign of damage.  Similar events were also observed in
the XMM-Newton EPIC MOS CCDs. The bright pixels and columns have been vetoed
on-board for the PC and WT modes. This is impossible in PD mode so that mode
is no longer used, which had the beneficial effect of reducing the CCD
temperature.

These two incidents had no direct impact on the spectroscopic performance of
the XRT, or its ability to image and locate new GRBs. Indeed, the XRT
routinely measures the early X-ray light-curves and spectra of all the GRB
afterglows at which it is promptly pointed (hence $95\%$ of the 183 GRBs
detected by the BAT for which a spacecraft slew was possible within 5 minutes
after the BAT trigger were detected by the XRT up until the middle of July 2007;
the BAT has detected a total of 243 GRBs up until the middle of July 2007). These
observations have revealed previously unexpected multiple
breaks~\cite{Zhang,Nousek} and flares in the early X-ray
light-curves~\cite{Zhang,Burrows07,King,Proga,Perna} suggesting extended
activity of the central engine up to $10^5$ s after the trigger for long and
short GRBs; which challenges the current progenitor models. Essential spectral
and temporal information was also obtained with the XRT for the peculiar event
GRB 060218 showing for the first time the rise of a
supernova~\cite{Campana06}\,. The XRT has also discovered the first short
burst afterglow~\cite{gehrels2} and provided accurate locations of several
short GRBs, which are associated with elliptical galaxies or star-forming
galaxies, the burst being located in the outskirts of the galaxy in the latter
case; these results tend to indicate that short GRBs are likely to be due to
binary compact object spiral-in and collision~\cite{Eichler,Rosswog}\,.  The
XRT also provides essential information for non-GRB targets, for example with
the follow-up of the recurrent nova RS Ophiuchi~\cite{Bode} or the
micro-quasar GRO J1655-40~\cite{Brocksopp}\,.  The fraction of time spent on
non-GRB targets (excluding the calibration targets) is $\sim 39.4\%$ since the
launch, and this is expected to increase over time.

In this paper, we concentrate on the XRT CCD in-flight spectral calibration,
and describe recent improvements to the response model made available as
response matrices through the HEASARC caldb (RMF version
009~\cite{Campana07}). The results obtained during the pre-flight calibration
were covered in Osborne et al.~\cite{Osborne}\,.





\section{IN-FLIGHT CALIBRATIONS} 

\begin{table}[h]
\caption{Summary of the in-flight calibration targets used up to July 2007.} 
\label{tab1}

\begin{center}       
\begin{tabular}{|l|l|l|l|l|} 
\hline
Object   & Type & Mode & Purpose &  Exposure (ks) \\
\hline
RXJ 1856.5-3754 & Neutron star & PC/WT & Low energy response & 55/47 \\
\hline 
PKS 0745-19 & Cluster of & PC & Effective area & 61 \\
            & galaxies   &    &                &     \\
\hline
2E 0102-7217 & SNR & PC & Gain, energy resolution and    & 76\\
             &     & WT & shoulder    & 55  \\
\hline
Cas A        & SNR & PC & Energy scale offset, gain, shoulder, &  144 \\
             &     & WT & CTI and energy resolution  &  51    \\
\hline
3C 273       & Quasar & WT & Effective area and cross-calibration& 18 \\
             &        &    & with XMM-Newton  &     \\
\hline
PSR 0540-69  & Pulsar & PC/WT & Effective area     & 56/26 \\
\hline
PKS 2155-304 & Blazar & WT & Effective area (cross-calibration & 13\\
             &      &    & with XMM-Newton)                   &\\
\hline
NGC 7172     & Seyfert 2 & PC & Redistribution & 15 \\
\hline
G21.5        & SNR    & PC & High-energy shelf  & 40 \\
\hline
Mkn 421      & Blazar & WT & Effective area and cross-calibration & 33 \\
             &        &    &  with XMM-Newton &    \\
\hline
Crab      & Pulsar & WT & Effective area  & 46 \\

\hline
\end{tabular}
\end{center}
\end{table} 


Since the FPCA front door was opened, spectroscopic calibrations have been
only performed using a set of well known celestial objects observed every six
months in order to monitor the change in the spectral response (see
Table~\ref{tab1}). The fraction of time spent on calibration targets is $\sim
7.5\%$ since the launch.  Many of our calibration targets are also used by
other X-ray observatories. This allows us to perform cross-calibration
campaigns with different X-ray instruments (e.g. the XMM-Newton EPIC MOS
cameras for the quasar 3C 273 and the blazar Mkn 421). In addition, we make
use of four $^{55}$Fe calibration sources permanently illuminating the non-imaging
area of the CCD. In order to optimise and facilitate the scheduling of the
calibration targets, a new state was implemented on-board the XRT in 2007 May.
This new state allows us, for instance, to control the observation mode while
the XRT is still in auto-state depending on the purpose of the observations.



\section{RESPONSE MODEL DEVELOPMENTS FROM JANUARY 2006 to JUNE 2007}

When an X-ray photon interacts within the CCD, it generates a charge cloud
which is collected in the depletion region after spreading in the bulk of the
detector. The charge cloud may spread into more than one pixel depending on
its energy and location of interaction. To compute the response matrices we
stack simulated spectra of monochromatic X-rays. We use the grade recognition
process used in the analysis software (for PC mode this is a $3\times 3$ pixel
matrix centred on the highest pixel). To avoid noise being included in the
charge summation and excessive telemetry usage, a threshold is set on-board
below which pixels are not considered.

The X-ray spectrum resulting from monochromatic radiation significantly
differs from a simple Gaussian, it consists of six components: a Gaussian peak
with a shoulder on the low energy side of the peak, an escape peak and a Si
K$\alpha$ fluorescence peak if the photon energy is above the Si K-shell edge,
a shelf extending to low energies, and at the very lowest energies a noise
peak (of which only the high energy side may be seen above threshold).

Below 1.5 keV, the shoulder and shelf are mainly produced by the charge losses
at the interface between the SiO$_2$ layer and the active silicon volume of
the open electrode, possibly due to a local inversion of the electric field
near the detector surface.  Above 1.5 keV, the shoulder and the shelf are
produced by several processes: (i) sub-threshold losses; (ii) recombination
and trapping in the bulk of the detector; (iii) inhomogeneity of the electric
field in the depletion depth, in addition to the surface losses.  Their exact
shapes depend on the readout mode, and hence they are different for the PC and
WT modes.  Both above and below 1.5 keV, the main peak broadens with time due
to the degradation of the charge transfer efficiency, producing a larger
shoulder at lower energies (see Section~\ref{CTI}).

Changes in our spectral response code have been made in order to better
reproduce the different components mentioned above.  The recent improvements
described below have been released as new spectral response files (version
v009~\cite{Campana07}).

\subsection{The shelf from photons above $\sim 2$ keV} 
\label{sec:shelf}

\begin{figure}[h]
\begin{center}
\begin{tabular}{cc}
\includegraphics[height=5.8cm]{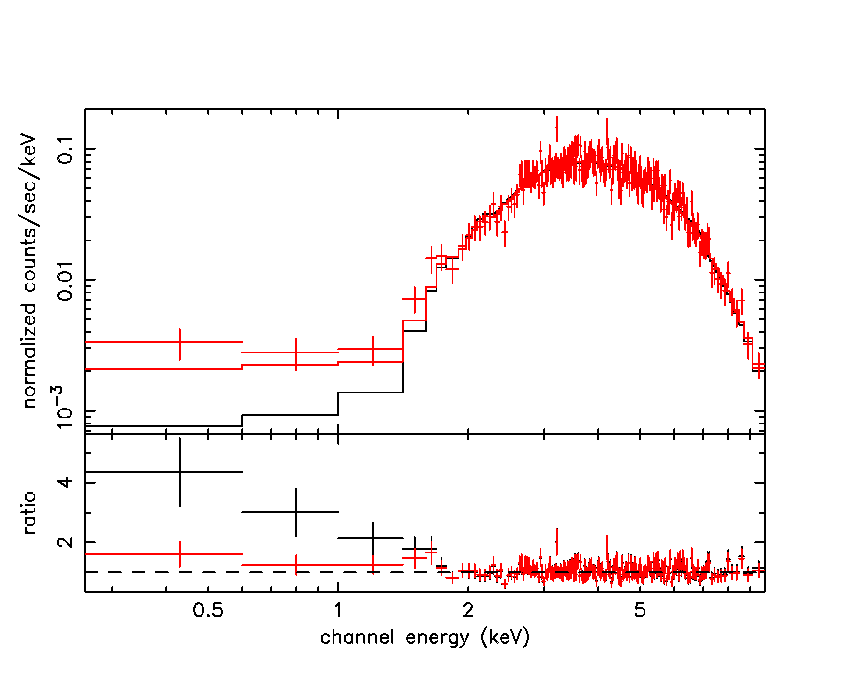} & \includegraphics[height=5.8cm]{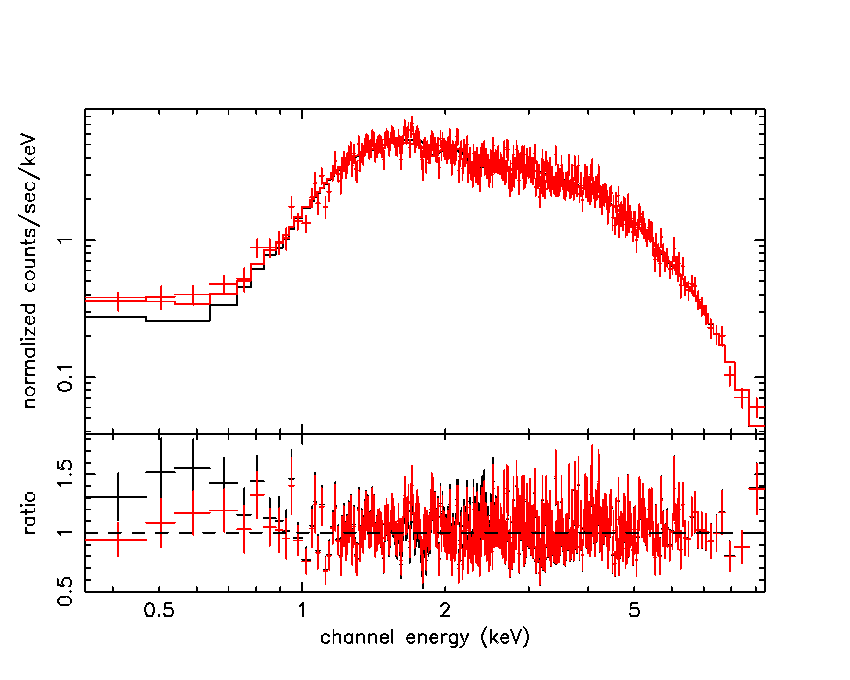}\\
\end{tabular}
\end{center}
\caption[]{Left:
  XRT PC grade 0-12 spectrum of NGC 7172 using the new v009 (red) and previous
  v008 (black) response files. The use of a {\scriptsize WABS*POWERLAW} model
  gives a value of $N_H \sim 7.3 \times 10^{22}$ cm$^{-2}$. Right: XRT
  WT grade 0-2 spectrum of the X-ray binary 4U 1608 using the new v009 (red)
  and previous v008 (black) response files. The use of a {\scriptsize
    WABS*(POWERLAW+DISKBB)} model gives a value of $N_H \sim 10^{22}$
  cm$^{-2}$.}
\label{fig_shelf}
\end{figure}

The new response matrice files (v009 RMFs) include an empirical rescaling of
the low energy shelf made by X-rays above 2\,keV, which significantly improve
the quality of spectral fits to the calibration sources (see
Fig.~\ref{fig_shelf}). This was previously incompletely modelled for either PC
or WT modes, resulting in an underestimation of the modelled redistributed
counts when fitting spectra of heavily absorbed sources (e.g. $N_H \ge
10^{22}$ cm$^{-2}$).

\subsection{The shoulder from photons above $\sim 1.5$ keV}

\begin{figure}[h]
\begin{center}
\begin{tabular}{cc}
\includegraphics[height=4.4cm]{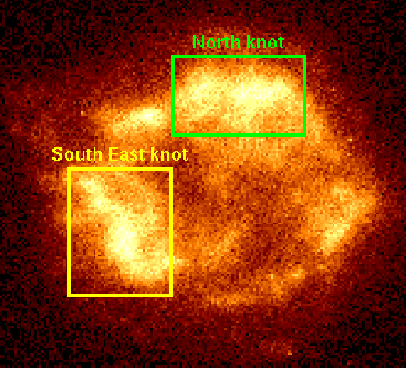} & \includegraphics[height=4.8cm,angle=0]{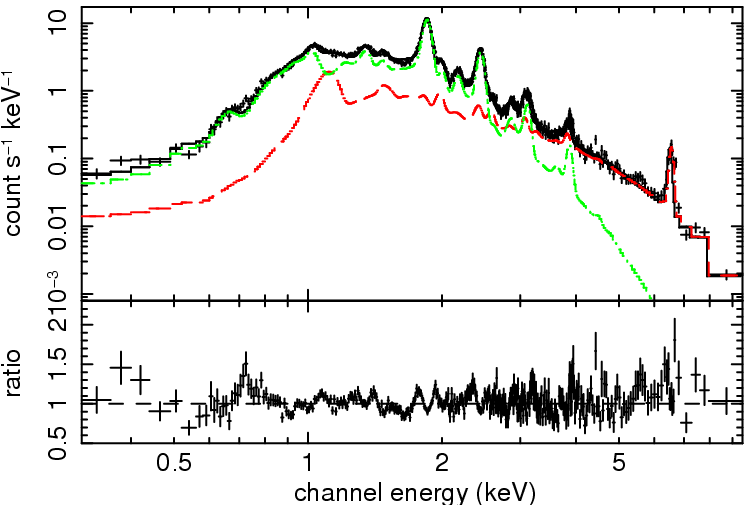}\\
\end{tabular}
\end{center}
\caption[]{Left: PC grade 0 image of the SNR Cas A. Right: PC grade 0 spectrum
  of the North knot of the SNR Cas A (see the image on the left). The data
  were taken in 2005 February 17. The spectrum is fit using a {\scriptsize
  PHABS(VNEI+VNEI)} model (see Willingale et al.~\cite{Willingale}). }
\label{fig_shoulder}
\end{figure}

Before the v009 release of the calibration files, the shoulder, which was
modelled by artificially increasing the event split threshold in order to
increase the sub-threshold losses, was not reproduced well (see Fig. 6a in
Osborne et al.~\cite{Osborne}).  We showed that modification of the shape of
the charge cloud formed in the field-free region using the formalism described
in Pavlov \& Nousek~\cite{Pavlov} can lead to a more physical modelling of the
shoulder (see Fig. 6b in Osborne et al.~\cite{Osborne}), the shape being no
longer a 2-D Gaussian (although a 2-D Gaussian remains sufficient in the
depletion region).  Fig.~\ref{fig_shoulder} shows the relatively good
agreement of the model with the very complicated spectrum of the SNR Cas A
(the North knot as noted in Fig.~\ref{fig_shoulder}).

\section{THE RESIDUALS AROUND AND BELOW THE OXYGEN EDGE}
\label{residual}

The fits of several continuum sources reveal negative (less than 20\%)
residuals around the Oxygen edge (0.54 eV) using v008 and v009 RMFs.

\begin{figure}[h]
\begin{center}
\begin{tabular}{cc}
\includegraphics[height=5.6cm]{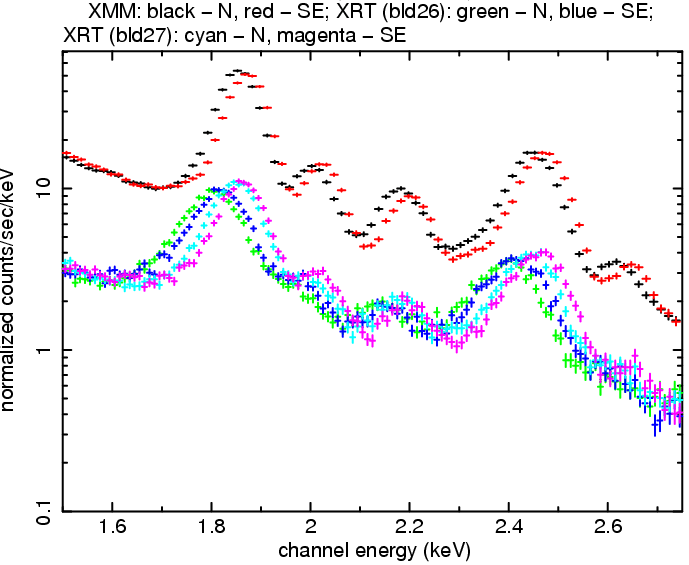} & \includegraphics[height=5.3cm]{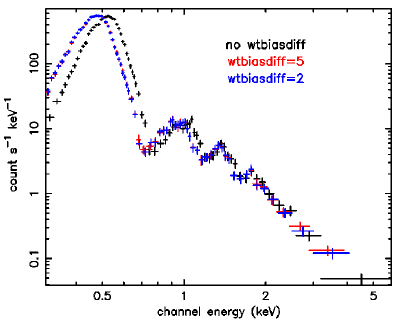}\\
\end{tabular}
\end{center}
\caption[]{Left: Comparison of the energy centroid of the Si and S lines in
  the North (N) and South East (SE) knots of the SNR Cas A as observed by the
  XMM MOS cameras (N: black; SE: red) and the XRT (N: green, cyan; SE: blue,
  magenta). The green and blue crosses correspond to XRT/PC grade 0 data for
  which the bias was contaminated by optical light from the sunlit Earth. The
  data were processed with the CALDB2.6 software which does not allow to bias
  corrections. In this case, an energy scale offset is observed when compared
  with the XMM MOS curves. The cyan and magenta crosses correspond to the same
  data processed with the new CALDB2.7 software including the tool
  {\scriptsize XRTPCBIAS}, which corrects the data.  Right: XRT WT grade 0-2
  spectrum of RS Ophiuchi: ({\it black}) the data not corrected to the bias
  problem and ({\it red and blue}) the data corrected using the new command
  option {\scriptsize WTBIASDIFF}.}
\label{fig_scale}
\end{figure}

Recently, observational evidence has shown that the CCD bias level can
significantly vary during the timescale of an orbital snapshot on an
astrophysical target. The bias level is mode-dependent and is subtracted
on-board during the XRT observations. Bias variations during a snap-shot can
occur due to changes in the CCD temperature and/or scattered optical light
from the sunlit Earth\cite{Beardmore}\,. This can result in energy scale
offsets, which give rise to residuals when fitting spectra, especially at low
energies (e.g.  around the Oxygen edge).  Energy scale offsets can be seen in
both PC and WT modes. A new command option was implemented in the 2.6 XRT
software ({\scriptsize WTBIASDIFF}) in order to correct the WT data for this
effect. A tool {\scriptsize XRTPCBIAS} included in the 2.7 XRT
software\,\footnote{see the following URL:
  http://heasarc.gsfc.nasa.gov/docs/software/lheasoft/release$_{-}$notes.html}
has been released in July 2007 by the XRT software team, which corrects the PC
data. The use of these contemporary time-dependent bias estimators can
significantly improve the energies of low energy events (see
Fig.~\ref{fig_scale}).

Charge transfer inefficiency due to accumulating proton damage also results in
a slow change in the energy scale. This is discussed in Section~\ref{CTI}.

Cross-calibration performed with other X-ray instruments in orbit such as {\it
  XMM-Newton} and {\it Suzaku} on the supernova remnant 2E 0102-7217 reveals
that at least these CCD cameras seem to suffer from an overestimation of the
model with respect to the data around the Oxygen edge.  Recently, the MOS
calibration team has decided to apply an {\it ad hoc} correction to their
quantum efficiency (QE) by decreasing the QE below the Oxygen edge by 10-15\%,
in order to correct the residuals.  We are investigating whether a similar
approach could work in the case of the XRT, once the effects of energy scale
offset have been corrected.

\section{CHARGE TRANSFER INEFFICIENCY}
\label{CTI}

\begin{figure}[h]
\begin{center}
\begin{tabular}{cc}
\includegraphics[height=5.8cm]{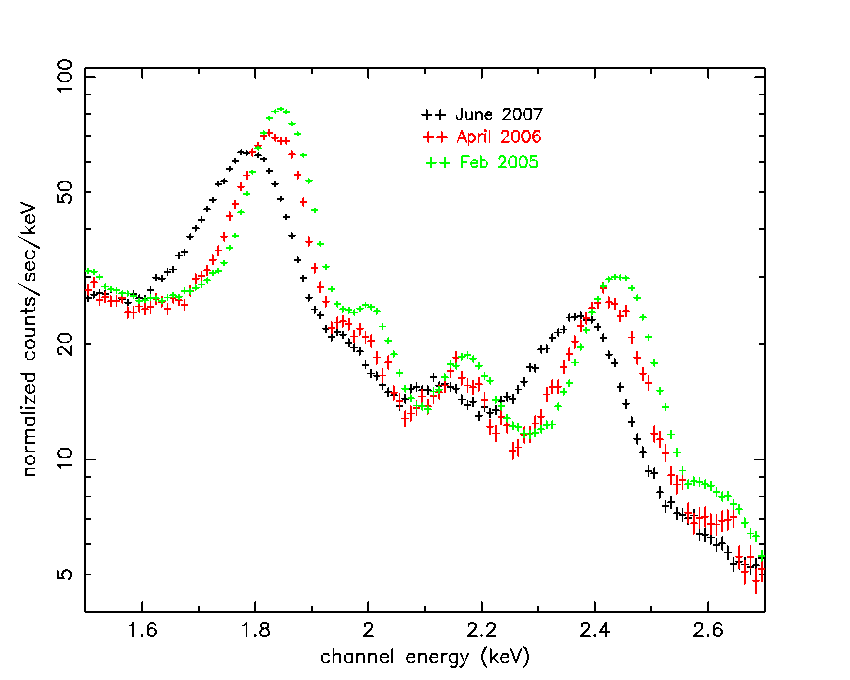}&\includegraphics[height=6.3cm,width=9cm]{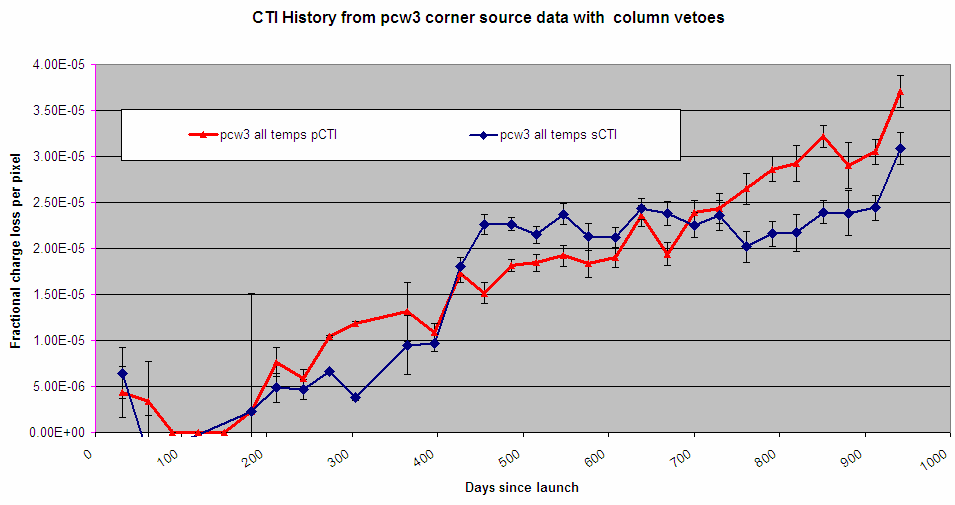}\\
\end{tabular}
\end{center}
\caption[]{Left: WT grade 0-2 spectrum of the SNR Cas A in the energy band of
  the Si and S lines at different epochs since the launch showing the
  degradation of the energy resolution and the gain due to the increase of
  traps in the CCD. The WT data were processed with the command option
  {\scriptsize WTBIASDIFF} correcting the bias level if corrupted (see
  Section~\ref{residual}). The measurements of the energy centroid of the Si
  and S lines imply an energy shift of about -50\,eV from February 2005 to
  June 2007, when the increase of CTI over time is not taking into account in
  the gain file.  The FWHM of the Si line increases from 105\,eV in February
  2005 to about $131$\,eV in June 2007.  Right: Evolution of the serial (blue)
  and parallel (red) CTI with time since the launch as measured using the four
  corner source data. The CTI values were computed vetoing the columns with
  significant traps.}
\label{fig_CTI}
\end{figure}

CCD detectors provide good X-ray imaging and spectroscopic performance.
However, the CCD energy resolution and gain degrade with time due to the
increase of the CTI. The main origin of CTI is the increase of charge traps,
which are mainly due to the irradiation of high-energy protons on the CCD
passing through the shielding. Although the low-Earth orbit of {\it Swift} and
the thick Al shielding around the CCD detector reduce the proton flux, the
frequent passages of the spacecraft through the South Atlantic Anomaly (SAA)
can cause formation of charge traps, and hence an increase of CTI. Since the
launch, the FWHM measured using the four $^{55}$Fe calibration sources
(located in each corner of the CCD; the area of the detector covered by these
corner sources being small and outside the imaging area) increased from 146 eV
at 5.9 keV in Feb 2005 to 210 eV in March 2007 when using the {\it bad and
  good columns} (i.e. columns with and without significant traps),
respectively. The broadening of the line is due to the energy scale shifting
effect of traps in the pixels through which the charge has to be transported.
The increase of traps in the CCD imaging area, the serial register and the
frame-store area also causes energy scale offset as shown in the left panel in
Fig.~\ref{fig_CTI}. In this Figure, the data were processed with a gain file
not corrected from the CTI increase.

Since we do not have any on-board calibration source illuminating the imaging
area, we use four $^{55}$Fe (5.9 keV) calibration sources to track the
temporal evolution of CTI. We define a parallel and serial CTI as follows: 
$$\mathrm{CTI}_{p,s} = \frac{E_i-E_{p,s}}{n\times E_i}$$ where
$\mathrm{CTI}_{p,s}$ corresponds to the parallel and serial CTI respectively,
and $E_i$ is the measured energy centroid of the corner source CS3 closest to
the output amplifier (bottom left of the imaging area), which does not suffer
from the CTI loss in the imaging area.  $E_{p,s}$ is the measured energy
centroid of the source located in the top left and bottom right corner of the
imaging area, respectively.  $n$ is the number of pixels that the charge has
to pass through in being readout (i.e.  transfered down the CCD and then along
the readout strip).  Note that we used only the columns with no significant
traps to do the computation, hence the CTI values derived are free from the
effects of the traps.  The computed parallel and serial CTI values (see the
right panel Fig.~\ref{fig_CTI}) as well as the gain values of the corner
source CS3 closest to the output amplifier are used to generate an
epoch-dependent gain file. The gain variations measured from the corner source
CS3 are due to the CTI increase over time in the store-frame area (in which no
CTI measurements can be performed since no calibration sources illuminate this
part of the CCD) and/or possible gain variations of the output FET (Field
Effect Transistor) with time. Up until July 2007, the gain of the corner
source CS3 increased by about $1\%$. While the parallel CTI increased
relatively steadily since the launch from $5\times 10^{-6}$ to $3.5\times
10^{-5}$ in June 2007 (see Fig.~\ref{fig_CTI}), the serial CTI increased
rapidly in a short period of time around November 2005 and stayed relatively
constant around $2.5\times 10^{-5}$ from December 2005 up to March 2007. An
updated v007 gain file in PC and WT modes will be soon released. 


In order to correct the data from the effects of traps, we plan to determine
the location and depth of the largest traps and implement a correction in the
data processing software.  We need to use a stable, sufficiently extended and
emission line-dominated source to map the central imaging area (a $200\times
600$ pixel window in PC mode and a 200 pixel window in WT).  Supernova
remnants (SNRs) are the best astrophysical sources.  We plan to use the SNR
Cas A which has intense Silicon and Sulphur lines. Although the energy
centroids show an energy variation of less than $\pm 10$ eV (at the energy of
the Si and S lines) across the remnant~\cite{Willingale}\,, the energy
variation due to traps that we want to measure are much larger.  The only
drawback of this method is that we cannot map all the imaging area ($600\times
602$ pixels) due to the excessive exposure time that would be needed.

A more efficient and simpler approach would be to use the charge injection
technique~\cite{Prigozhin} to measure the CTI for each column, as it has been
done for the Soft X-ray cameras on-board {\it Suzaku}~\cite{Nakajima}\,, since
we can then correct all the CCD imaging area without long calibration
exposures. We are undertaking further work to verify the feasibility of this
technique in the case of the e2v CCD-22 used in the XRT.

\section{REDUCTION OF NOISE BY SUBSTRATE VOLTAGE CHANGE}

The loss of the active cooling causes the XRT to operate at higher than
expected temperatures (see Section 1); this generates significant
thermally-induced noise appearing as low energy events. Raising the substrate
voltage as described in Osborne et al.~\cite{Osborne} will allow a reduction
of this noise and the use of lower energy X-ray events, since the volume of
Silicon in which carriers are generated is reduced.

\begin{figure}[h]
\begin{center}
\begin{tabular}{cc}
\includegraphics[height=5.3cm]{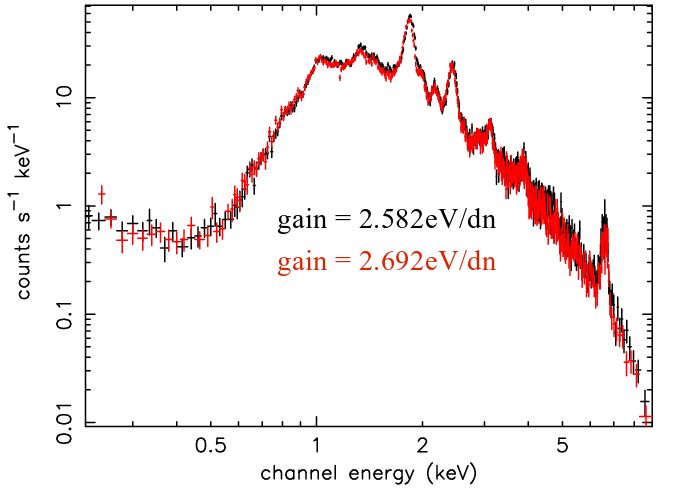} & \includegraphics[height=5.8cm]{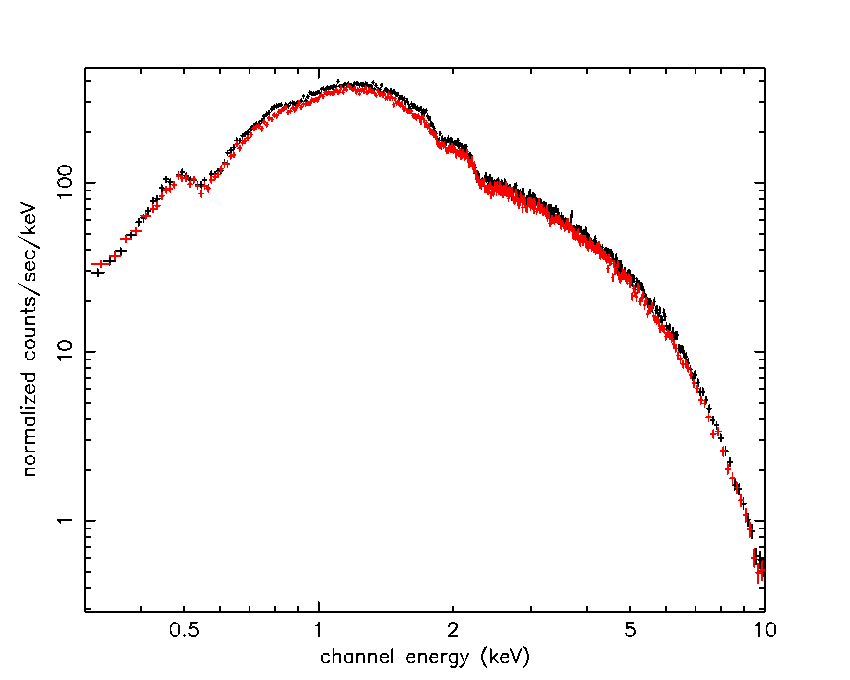}\\
\end{tabular}
\end{center}
\caption[]{Left: WT grade 0-2 spectra of Cas A with the substrate voltage
  $V_{\rm ss} = 6$ V (red crosses) and $V_{\rm ss} = 0$ V (black crosses). Note that
  the gain changed between the two configurations, since the gain of the
  output FET slightly changed between the two configurations (see
  text).  Right: WT grade 0-2 off-pulse spectra of the Crab with $V_{\rm ss} = 6$
  V (red crosses) and $V_{\rm ss} = 0$ V (black crosses). The two plots show that
  the changes in the effective area, hence the QE, are minimal between the two
  configurations.  }
\label{fig_substrate}
\end{figure}

Osborne et al.~\cite{Osborne} have shown that raising the substrate voltage
will cause a decrease of the depletion depth, resulting in a migration of
events to higher grades (hence, an increase of the sub-threshold losses) and a
slight decrease of the QE. We performed observations of Cas A (2.2\,ks in PC
and 0.8\,ks in WT) and the Crab (0.6\,ks in WT) with raised substrate voltage
($V_{\rm ss} = 6$ V instead of 0 V).  Fig.~\ref{fig_substrate} shows that the
changes in the effective area are small between $V_{\rm ss} = 6$ V and 0 V
(i.e.  less than 15\% at 6 keV).  The gain will also change when the substrate
voltage settings are modified, because the gain of the output FET, which works
in a source follower configuration, will slightly change in this new
configuration.

\section{CONCLUSION}

We showed that our Monte-Carlo code computing the PC and WT RMFs allows us to
describe the CCD response well. Thus, the v009 RMF files release in July 2007
allow a better modelling of the low-energy shelf of heavily absorbed sources
and a better physical description of the origin of the shoulder from photons
above 1.5 keV. We also showed that energy scale offsets due to corruption of
the bias level in both PC and WT modes can limit the improvements of the
instrument spectral response, especially at low energies. New tool
{\scriptsize XRTPCBIAS} and the new command option {\scriptsize WTBIASDIFF}
have been provided by the XRT software team to correct the bias level in both
PC and WT mode, and hence suppress energy scale offset. These tools are
included in the 2.7 XRT software package recently released.
   
We are working on a more complete trap characterisation. To do so, the XRT
began observing the SNR Cas A for 80\,ks in PC mode and 50\,ks for WT. Because
of the development of traps with time in the CCD detector, we also outline the
importance to use an on-board calibration source which fully illuminates the
focal plane, for instance as used on-board XMM-Newton, for future X-ray CCD
instruments.

Finally, in order to improve further the XRT spectral performance, we plan
to raise permanently the substrate voltage to $V_{\rm ss}=6$\,V on-board before
the end of 2007.  An intense phase of re-calibration of the instrument will
shortly follow to update the spectral response files, the gain file and the
ground data processing software.


\subsection{Acknowledgements} 
            
OG, APB, AFA, KLP, JPO, LT, PE, RS gratefully acknowledge PPARC funding. This
work is supported in Italy by ASI grant I/R/039/04 and the Ministry of
Universiy and Research of Italy (PRIN 2005025417), and at Penn State by NASA
contract NAS5-0136.


\end{document}